\documentclass[aps,pra,amsmath,amssymb,reprint,showpacs]{revtex4-1}

\usepackage{graphicx}
\usepackage{dcolumn}
\usepackage{bm}

\DeclareMathOperator\erfc{erfc}

\begin{document}

\title{Cold beam of isotopically pure Yb atoms by deflection using 1D-optical molasses}

 \author{Ketan D. Rathod}
 \affiliation{Department of Physics, Indian Institute of
 Science, Bangalore 560\,012, India}
 \author{P. K. Singh}
 \affiliation{Department of Physics, Indian Institute of
 Science, Bangalore 560\,012, India}
 \author{Vasant Natarajan}
 \affiliation{Department of Physics, Indian Institute of
 Science, Bangalore 560\,012, India}
 \email{vasant@physics.iisc.ernet.in}
 \homepage{www.physics.iisc.ernet.in/~vasant}

\begin{abstract}
We demonstrate generation of an isotopically pure beam of laser-cooled Yb atoms by deflection using $1$D-optical molasses. Atoms in a collimated thermal beam are
first slowed using a Zeeman Slower. They are then
subjected to a pair of molasses beams inclined at $45^\circ$ with respect to the slowed atomic beam. The slowed atoms are deflected and probed at a distance of $160$ mm. We demonstrate selective deflection of the bosonic isotope $^{174}$Yb, and the fermionic isotope $^{171}$Yb. Using a transient measurement after the molasses beams are turned on, we find a longitudinal temperature of $41$ mK.
\end{abstract}



\maketitle

\section{Introduction}
Cold atoms \cite{JOSAB89}, with their long measurement times, promise to revolutionize the field of precision measurements. In this respect, laser-cooled Yb constitutes a useful species because its spin-zero ground state obviates the need for a second repumping laser, as is required for laser cooling of the more common spin-half alkali-metal atoms. As a consequence, cold Yb has been proposed for use in next-generation atomic clocks \cite{HZB89,HBO05}, and the search for a permanent electric dipole moment (EDM) \cite{NAT05}. Existence of an EDM is an indication of both parity violation (which is already known from the weak interaction) and time-reversal symmetry violation in the fundamental laws of physics, and is one of the most important experiments in atomic physics today. As a consequence EDM measurements have been reported in $^{199}$Hg \cite{GSL09}, $^{133}$Cs \cite{MKL89}, $^{205}$Tl \cite{RCS02}, TlF \cite{CSH91}, $^{174}$YbF \cite{HST02}, etc. Both clock and EDM measurements gain from having a cold {\em continuous} beam of atoms that is separate from the cooling laser beams. For atomic clocks, a continuous beam avoids intermodulation or the Dick effect \cite{GDT07}, seen in pulsed fountain clocks. For EDM experiments, the electric-field plates can be brought very close because there is no interference from any laser beams.

Here, we demonstrate such a continuous beam by deflection using a pair of $1$D-optical molasses beams. Atoms emanating from a thermal source are first cooled and slowed in a Zeeman slower. The molasses beams are inclined at $45^\circ$, and chosen to be nearly resonant with one particular isotope. Thus, the deflected atomic beam is isotopically pure, and free from both other isotopes and unslowed atoms.  The deflected atoms are probed at a distance of $160$ mm from the molasses region. We verify that any selected isotope can be deflected: $^{174}$Yb as an example of an even isotope, and $^{171}$Yb as an example of an odd isotope. A transient measurement after the molasses beams are turned on shows a mean velocity of $15.55$ m/s, and a longitudinal temperature of $41$ mK. This temperature represents a factor of three improvement over our recent work in which atoms were launched vertically from a two dimensional magneto-optic trap ($2$D-MOT) \cite{RSN13}. Compared to the earlier experiment, the current set-up has the additional advantage of being easier to implement.

\section{Experimental details}

\begin{figure}
\begin{center}
\includegraphics[width=0.95 \columnwidth]{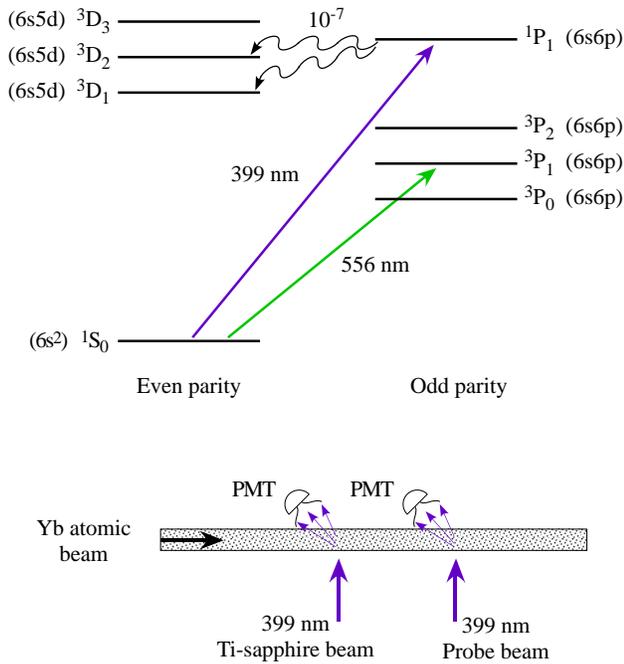}
\caption{(Color online) (a) Low-lying energy levels of Yb. Even parity levels are shown on the left and odd parity levels on the right. Therefore, only transitions between the left manifold and right manifold are electric dipole allowed. (b) Spectroscopy arrangement used to determine the frequency of the lasers.}
\label{levels}
\end{center}
\end{figure}

Many of the experimental details are similar to our earlier work in Ref.\ \cite{RSN13}, and are presented here for completeness. Yb has two cooling transitions---the strongly-allowed $^1{S_0} \rightarrow {^1P}_1$ transition at $399$~nm, and the weakly-allowed $^1{S_0} \rightarrow {^3P}_1$  intercombination line at $556$~nm---as shown in Fig.\ \ref{levels}. In this study (as in our previous work), we have only used the former one. Though its relatively large linewidth of $28$~MHz implies a large Doppler-cooling temperature of $690$~$\mu$K, it allows for Zeeman slowing over a short distance. The saturation intensity for this transition is $58$ mW/cm$^2$. Two of the nearby triplet-$D$ states are lower in energy than the ${^1P}_1$ state, so the transition is not really closed, but the branching ratio of $10^{-7}$ to these states is negligibly small \cite{HTK99}. For example, the number of photons required to slow an atom from $300$ m/s to $25$ m/s is less than $10^5$. The two isotopes used in this study are $^{174}$Yb (boson with $I=0$), and $^{171}$Yb (fermion with $I=1/2$). Therefore, the isotope $^{174}$Yb has a single hyperfine transition from $F=0 \rightarrow F'=1$. The other isotope $^{171}$Yb has two transitions: $F=1/2 \rightarrow F'=1/2$ and $F=1/2 \rightarrow F'=3/2$. The Zeeman shifts for the $0 \rightarrow 1$ transition in $^{174}$Yb and the $1/2 \rightarrow 3/2$ transition in $^{171}$Yb both have the same value of $1.4$ MHz/G, hence the same Zeeman-slower profile and the slower-beam detuning can be used for both isotopes.

The main laser for accessing the $399$ nm transition is generated in a two-step process. We start with a {\em single-frequency} Ti:sapphire laser (Coherent $899-21$) operating at $798$ nm, pumped with $532$ nm light (Spectra Physics Millennia X). Its output is frequency doubled to $399$ nm in an external-cavity doubler with a lithium triborate crystal (Laser Analytical Systems), with conversion efficiency of about $12$\%. The output of the doubler is split into three parts: the first part for the Zeeman-slowing beam, the second part for the $1$D-molasses beams, and the third part to monitor its frequency in the fluorescence spectroscopy setup shown in the bottom of Fig.\ \ref{levels}. The fluorescence light is collected using a Hamamatsu R$928$ photomultiplier tube (PMT). The laser frequency is manually adjusted to be at the fluorescence peak, and left there for the duration of the experiment. The drift of the Ti:sapphire laser is small enough that there is no significant movement away from the peak. The frequency shift required for the Zeeman-slowing and molasses beams are produced using acousto-optic modulators (AOMs). The deflected atoms are probed using a second low-power laser, composed of a grating-stabilized diode laser (Nichia Corporation). Its frequency is also monitored in the same spectroscopy set up using a second PMT.

\begin{figure}
\begin{center}
\includegraphics[width=0.95 \columnwidth]{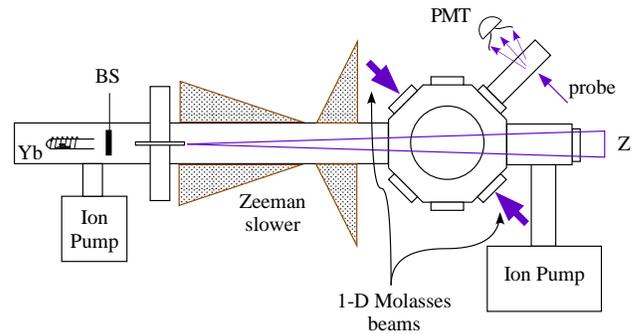}
\caption{(Color online) Top view schematic of the vacuum chamber used for the experiment. Detailed description is given in the text. Z: Zeeman-slower beam, BS: Beam shutter.}
\label{schema}
\end{center}
\end{figure}

A top-view schematic of the vacuum system used in the experiment is shown in Fig.\ \ref{schema}. The source of atoms is a resistively-heated quartz ampoule containing all the isotopes of Yb in their natural abundance. The source region is maintained at a pressure below $10^{-7}$ torr using an ion pump of $20$ l/s speed. This region is attached to the experimental chamber through a differential-pumping tube, so as to allow for a pressure difference of 2 orders of magnitude. The first part is a Zeeman-slowing region, consisting of a stainless-steel (SS) tube with OD $42$ mm and length $500$ mm. The second part is the main experimental chamber, consisting of an octagonal SS cell in the horizontal plane with two $70$ mm viewports for the $1$D-molasses beams. This region has a pressure of $10^{-9}$ torr. The port at $45^\circ$ with respect to the direction of the atomic beam has a rectangular glass cell for probing the deflected atoms. The probe region is at a distance of $160$ mm from the deflection point, and the fluorescence signal is collected with a third PMT. The entire system on this side of the differential pumping tube is pumped by a second ion pump with $55$ l/s speed.

Atoms emanating from the oven are first slowed in a spin-flip Zeeman slower \cite{PAN10}, i.e.\ one with a positive initial field and a negative final field going through zero in between. The main advantage of this design is that the field coils do not carry much current, and do not require active cooling. With a detuning of $-420$ MHz and an initial field of $+290$ G, the slower is designed to slow all atoms with a velocity less than $330$ m/s. This corresponds to capturing $55$\% of the atoms emanating from the oven when it is heated to $400^\circ$C. The final field at the end of the $0.33$ m long slower is $-260$ G, which means that the velocity after the slower is $23$ m/s. The coils required for generating the slower-field profile are made by winding welding wire around the outside of the Zeeman-slower tube. The wire carries $34$ A for the forward slower, and $26$ A for the reverse part. The slowing beam is circular, with $1/e^2$ diameter of $30$ mm at the entrance to the octagonal chamber, and focussed to a spot at the end of the differential-pumping tube using a convex lens of focal length $1$ m.

The incoming molasses beam has a total power of $46$ mW. It is elliptic in cross section, with $1/e^2$ diameter of $10 \times 15$ mm, and the long axis aligned in the vertical direction. Therefore, the peak intensity at the beam center is $78$ mW/cm$^2$, to be compared to the saturation intensity of $58$ mW/cm$^2$. The beam has a detuning of $-14$ MHz ($=-\Gamma/2$ for this transition), which is known to give the lowest temperature for optical molasses \cite{LPR89}. It is linearly polarized. The return beam is generated by retro-reflecting the incoming beam with a plane mirror. Since the velocity of atoms after the slower is $23$ m/s, and the molasses beams are inclined at an angle of $45^\circ$, the expected mean velocity of the deflected atoms is $23 / \sqrt{2} = 16.26$ m/s.

The experimental parameters used in this study are summarized in Table \ref{pars}.

\begin{table}
\caption{Experimental parameters used for the experiment.}
\begin{center}
\begin{ruledtabular}
\begin{tabular}{ll}
Zeeman-slower beam power & $10$ mW\\
Zeeman-slower beam detuning  & $-420$ MHz\\
1D-molasses beam intensity (max) & $78$ mW/cm$^2$\\
1D-molasses beam detuning & $-14$ MHz 
 \label{pars}
\end{tabular}
\end{ruledtabular}
\end{center}
\end{table}

\section{Results and discussion}

\begin{figure}
\begin{center}
\includegraphics[width=0.95 \columnwidth]{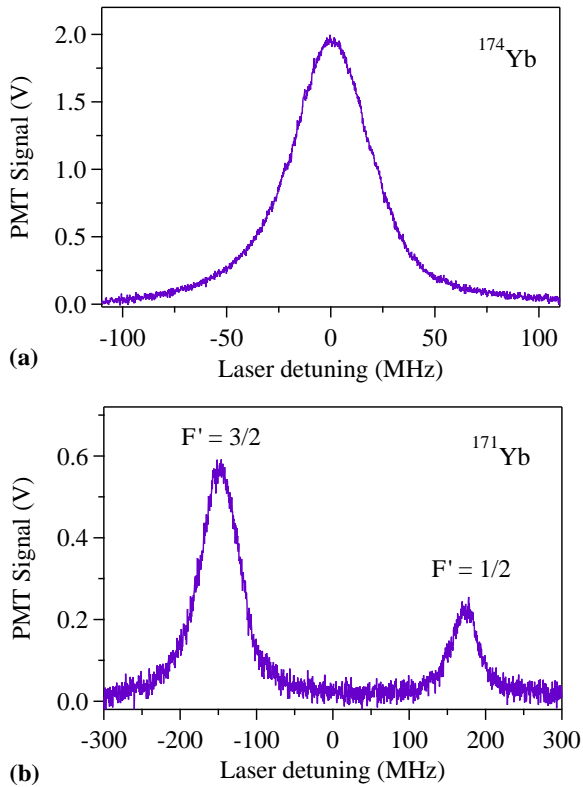}
\caption{(Color online) Fluorescence signal from deflected atoms in the probe region of (a) $^{174}$Yb and (b) $^{171}$Yb. There are two peaks for $^{171}$Yb because it has two hyperfine levels in the excited state, with $F'$ values as labeled.}
\label{cw}
\end{center}
\end{figure}

The first experiment was to make sure that the deflected atoms were isotopically pure. The deflected isotope was either  $^{174}$Yb (as an example of a boson), or $^{171}$Yb (as an example of a fermion). The main Ti-Sa laser was brought into resonance with the desired isotope, and the probe laser was scanned across the same isotope. The PMT signal for the two isotopes is shown in Fig.\ \ref{cw}. The two peaks for $^{171}$Yb correspond to the $1/2 \rightarrow 3/2$ and $1/2 \rightarrow 1/2$ transitions.

There are two things to note about the spectra. One is that the linewidth of the peaks is $44$ MHz, which is $1.5$ times the natural linewidth of $28$ MHz. This is typical for such spectra seen in our previous work \cite{BRD03,DBB05}, and is due to power broadening and transverse temperature. The second point is that the peak height for $^{174}$Yb is $3.5$ times larger than the $1/2 \rightarrow 3/2$ peak in $^{171}$Yb, resulting in correspondingly higher signal-to-noise ratio (SNR). This is again similar to the ratio of these two peaks seen in our earlier work, and is due to a combination of the smaller natural abundance and hyperfine structure for $^{171}$Yb.
 
The second experiment was designed to
measure the longitudinal temperature in the deflected atomic beam. For this, we look at the fluorescence
signal as a function of time after the molasses beams are
turned on. The only isotope studied was $^{174}$Yb. The probe laser is now locked to this resonance peak. The measured signal is therefore determined by both the mean velocity, and the longitudinal spread around the mean or temperature. If the mean velocity is $\bar{v}$, and the distribution follows a Maxwell-Boltzmann curve at a temperature $T$, then the probability density function is
\begin{equation}
f(v) = \sqrt{\frac{m}{2 \pi k_B T}} \exp \left[ -\frac{m(v-\bar{v})^2}{2 k_B T} \right]
\, ,
 \label{dist}
\end{equation}
so that $f(v)\,dv$ is the probability of the atom having a velocity between $v$ and $v+dv$.

The measured signal is proportional to the total number of atoms in the probe region at a given time $t$. If the distance to this region from the deflection region is $d$, then all atoms having an initial velocity greater than $d/t$ will contribute to the signal. Therefore, the signal at a time $t$ is proportional to the integral of the above distribution from $d/t$ to $\infty$, i.e.\
\begin{equation}
N(t) = N_0 \erfc \left[ \sqrt{\frac{m}{2 k_B T}} \left({\frac{d}{t} - \bar{v}} \right)\right]
\,
 \label{erfc}
\end{equation}
We neglect the effect of gravity because the distance by which atoms fall in the time it takes to reach the probe region is about $0.5$ mm, which is much smaller than the probe-beam diameter.

\begin{figure}
\begin{center}
\includegraphics[width=0.95 \columnwidth]{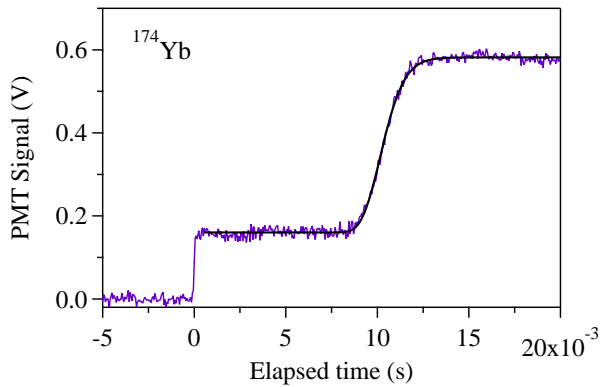}
\caption{(Color online) Fluorescence signal of $^{174}$Yb atoms in the probe region as a function of elapsed time after the $1$D-optical molasses beams are turned on. The probe laser is locked to the resonance peak. The sudden jump at $t=0$ is due to stray light from the molasses beams making it to the PMT. The smooth line is a curve fit the expected line shape [Eq.\ (\ref{erfc}) in the text].}
\label{fvst}
\end{center}
\end{figure}

The results of this experiment are shown in Fig.\ \ref{fvst}. The jump in signal at $t=0$ is because of stray light from the molasses beams making it to the PMT. The smooth line is a curve fit to the line shape given in Eq.\ (\ref{erfc}). The fit yields a mean velocity of $\bar{v} = 15.52(2)$ m/s, consistent with the expected value of $16.26$ m/s. The spread around the mean is $1.98(3)$ m/s. This corresponds to a longitudinal temperature of $41(1)$ mK, which represents an improvement of a factor of $3$ over our previous work in Ref.\ \cite{RSN13}.

\section{Conclusion}
In summary, we have demonstrated the creation of an isotopically pure cold beam of Yb atoms via deflection using $1$D-optical molasses. A continuous cold beam has many advantages over the more common pulsed fountain for precision measurements. Yb atoms emanating from a thermal source are first slowed in a Zeeman slower, then deflected using a pair of molasses beams inclined at $45^\circ$ with respect to the slowed atomic beam. The deflected atoms have a longitudinal temperature of $41$ mK, which is more than $3$ times better than our previous experiment \cite{RSN13}. The atoms in this study were deflected using molasses in one direction, and we hope to increase the flux by using an additional set of molasses beams orthogonal to the plane of octagonal cell and hence the atomic beam---a configuration called $2$D-optical molasses.

\begin{acknowledgments}
This work was supported by the Department of Science and Technology, India, through the Swarnajayanti fellowship. K.D.R. acknowledges financial support from the Council of Scientific and Industrial Research, India. P.K.S. acknowledges financial support from University Grants Commission, India.
\end{acknowledgments}


%

\end{document}